\documentclass[reprint,twocolumn,aps,pra,nofootinbib,groupedaddress,amsmath,amssymb,
longbibliography]{revtex4-2}
\usepackage{graphicx}
\usepackage{txfonts,comment}
\usepackage{bm}
\usepackage[colorlinks=true,	linkcolor=blue,urlcolor=blue,anchorcolor=blue,citecolor=blue,bookmarksnumbered]{hyperref}
\usepackage[mathscr]{euscript}
\usepackage[english]{babel}
\usepackage{natbib}

\def\be{\begin{equation}}
	\def\ee{\end{equation}}
\def\bea{\begin{eqnarray}}
	\def\eea{\end{eqnarray}}
\def\bes{\begin{subequations}}
	\def\ees{\end{subequations}}

\begin{document}
\title{Kinetic Slush in Dissipative Rydberg Arrays}
\author{Hufeng Wen$^{1}$}
\author{Ya-Xin Xiang$^{1}$}
\author{Qun-Li Lei$^{1, 2, 4}$}
\author{Thomas Pohl$^{3}$}
\author{Zhengyang Bai$^{1, 2}$}
\email{zhybai@nju.edu.cn}
\author{Yu-Qiang Ma$^{1, 2, 4}$}
\email{myqiang@nju.edu.cn}
\affiliation{
		$^1$National Laboratory of Solid State Microstructures and School of Physics, Collaborative Innovation Center of Advanced Microstructures, Nanjing University, Nanjing 210093, China\\
        $^2$Jiangsu Physical Science Research Center\\
        $^3$Institute for Theoretical Physics, TU Wien, Vienna 1040, Austria\\
		$^4$Hefei National Laboratory, Hefei 230088, China\\}
\date{\today}	
\begin{abstract}
We investigate the nonequilibrium dynamics of a kinetic slush state realized via facilitated excitations in a two-dimensional dissipative Rydberg atom array. Utilizing kinetic Monte Carlo simulations in the strong dephasing regime, we show that tuning the laser detuning to compensate for nearest-neighbor interactions enables facilitated dynamics, while residual next-nearest-neighbor interactions induce kinetic frustration. This competition gives rise to an emergent slush phase characterized by pronounced dynamical heterogeneity and the coexistence of liquid- and solid-like configurations. By probing its nonequilibrium real-time evolution, we uncover a dynamical crossover from isotropic spreading to one-dimensional constrained growth, which can be continuously tuned by the frustration strength. Within this crossover regime, the slush dynamics generate fractal cluster geometries and generalized Ornstein-Zernike spatial correlations.  Finite-size scaling analysis supports the system's proximity to criticality.  Our findings establish driven-dissipative Rydberg systems as a highly controllable platform for engineering complex nonequilibrium soft-matter phenomena.
\end{abstract}
\maketitle
\textit{Introduction}.--- 
Uncovering the nature of strongly correlated quantum matter in and out of equilibrium remains a foundational challenge, as the complex interplay of competing interactions drives the emergence of exotic quantum phases~\cite{Bloch_Many_RMP_2008, gross2017Quantum, browaeys2020Manybody, sieberer2025Universality, schauss2012Observation, weber2015Mesoscopic, helmrich2018Uncovering, serbyn2021Quantum, bluvstein2021Controlling, Semeghini_topo_2021, widmann2026Observationa}. Often, the most intriguing of these are complex composite phases that defy simple classification by combining traditionally distinct forms of ordering. A celebrated example is the supersolid, which combines crystalline order with superfluidity~\cite{Leggett_superSolid,Supersolid_Henkel_2010,leonard2017Supersolidb,Supersolidity_Zhang_2019,Supersolid_PRX_2019,casotti2024Vortices,trypogeorgos2025Emerginga}. Another paradigm of such mixed-order phenomena is the ``slush" state—a highly heterogeneous phase characterized by the real-space coexistence of fluid-like active regions and solid-like arrested backbones~\cite{rau2016Spin, zhang2024Quantum}.

Originally proposed in an extended classical spin-ice model, spin slush arises from competing interactions that reshape the ground-state manifold and support branching defect structures~\cite{rau2016Spin}.
A complementary route to exploring slush behavior is through kinetic constraints that restrict transitions between many-body configurations~\cite{zhang2024Quantum}. Rydberg systems offer a controllable setting for realizing such constrained dynamics~\cite{lesanovsky2013Kinetic,lesanovsky2014Outofequilibrium,valado2016Experimental,marcuzzi2017Facilitation}, with optical tweezer arrays providing control over atomic positions and excitation~\cite{labuhn2016Tunable, bernien2017Probing, zhang2017Observation, scholl2021Quantum, ebadi2021Quantum, Schlosser_PRL_2023, kim2024Realization, Cheng_Science_2025, manetsch2025tweezer, AI_2025_Lin}.
Their programmability enables the realization of the facilitation (antiblockade) mechanism, where the excitation of an atom is strongly enhanced in the vicinity of an already excited one~\cite{Antiblockade_ates_2007,Antiblockade_Amthor_2010,marcuzzi2017Facilitation,helmrich2020Signaturesa,wintermantel2021Epidemic,brady2024Anomalousa}.
Recently, a quantum analogue of slush was proposed in closed Rydberg arrays with a strict single-nearest-neighbor (NN) facilitation constraint, exhibiting loopless branching structures and algebraically decaying spatial correlations at zero temperature~\cite{zhang2024Quantum}.

While the equilibrium properties and spatial correlations of quantum slush have been explored~\cite{zhang2024Quantum}, its dynamical heterogeneity and nonequilibrium dynamics in Rydberg arrays remain to be understood. In practice, because the emergence of this state is inherently governed by long-time many-body dynamics, environmental dissipation inevitably dictates the system's prolonged evolution~\cite{deleseleuc2018Analysis,guardado-sanchez2018Probinga,levine2018HighFidelity,lesanovsky2013Kinetic,lesanovsky2014Outofequilibrium,marcuzzi2014Effective,valado2016Experimental,schempp2014Full,malossi2014Full,marcuzzi2017Facilitation,helmrich2020Signaturesa,QDCTC_2025,chen2025Collective,zhang2025observation}.
This environmental coupling affects both quantum coherence and the formation and relaxation of heterogeneous spatial structures. A central question is whether a slush state can emerge under these dissipative conditions and how its characteristic dynamical heterogeneity manifests during nonequilibrium evolution.

\begin{figure}[htbp]
		\centering
		\includegraphics[width=0.48\textwidth]{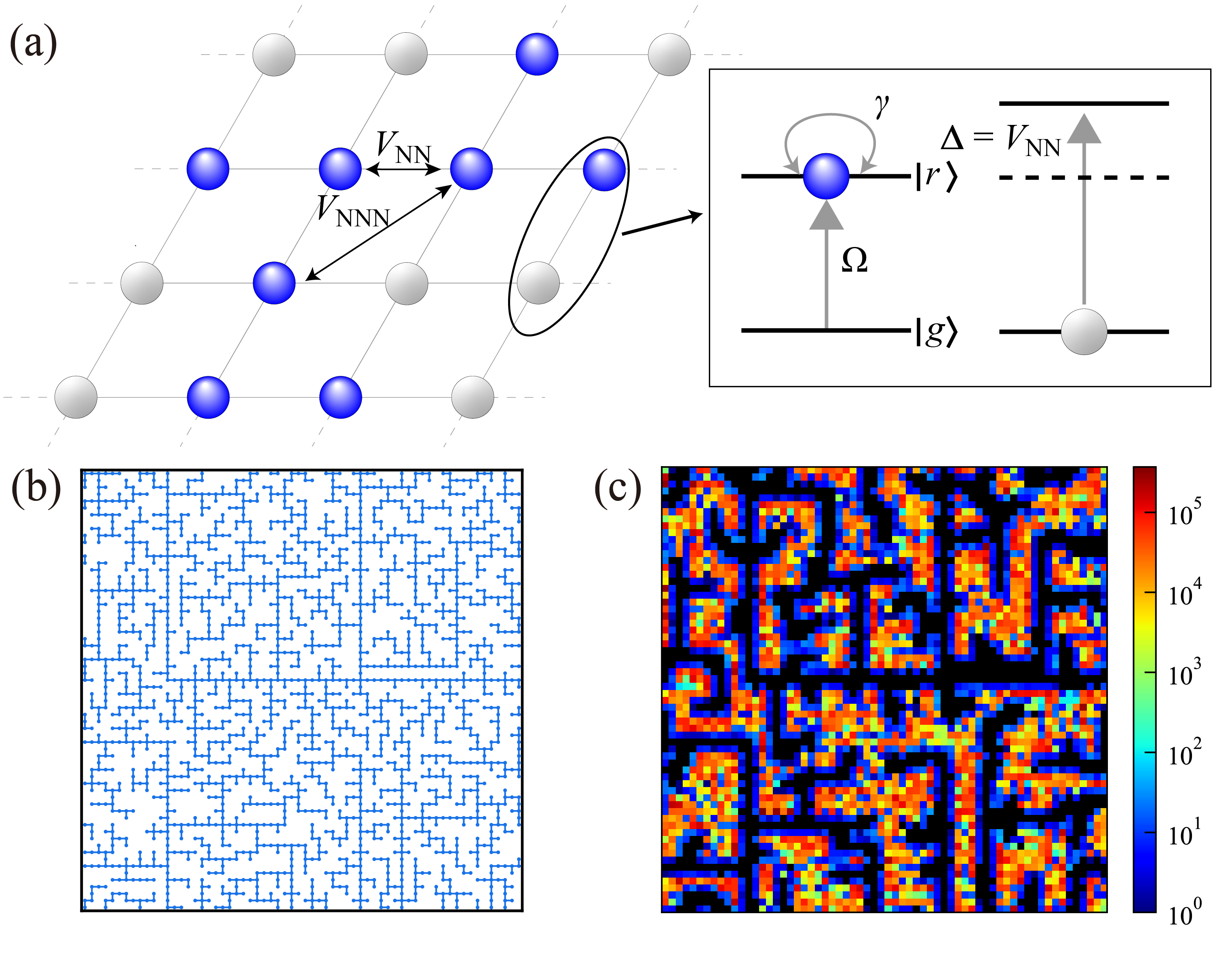}
		\caption{\label{fig1} 
\textbf{Model setup and frustrated Rydberg kinetics.} (a) Schematic of two-level atoms on a square lattice driven by a laser with Rabi frequency $\Omega$ and detuning $\Delta$. The condition $\Delta=V_{\mathrm{NN}}$ facilitates excitation exclusively when exactly one NN is in the Rydberg state $|r\rangle$. Under this condition,   $V_{\mathrm{NNN}}$ acts as a kinetic frustration that modulates the local spin-flip rate. (b) Snapshot of the steady-state spatial configuration. (c) Corresponding heatmap of local activity $\mathcal{K}_{j}$, measuring the number of spin flips at site $j$ over an observation window $T = 10^6$. Dark-blue regions indicate low-activity, solid-like sites within the observation window, while red/yellow regions denote highly active, liquid-like sites. Simulations in (b) and (c) are performed using the KMC method on a $64\times 64$ lattice with  $\eta =100$.}
	\end{figure}

In this work, we report on the nonequilibrium dynamics of a kinetic slush state realized in a two-dimensional (2D) dissipative Rydberg atom array [Figs.~\ref{fig1}(b) and (c)]. By tuning the laser to compensate for the NN interaction, we activate facilitated dynamics. We find that stabilizing this slush state hinges on the often-neglected van der Waals (vdW) tail: while NN interactions drive facilitation, next-nearest-neighbor (NNN) or higher-order couplings introduce the essential kinetic frustration. 
Employing kinetic Monte Carlo (KMC) simulations under strong dephasing~\cite{ruizbarlett2009Differences,Antiblockade_ates_2007,lesanovsky2013Kinetic,sibalic2016Drivendissipative,brady2024Anomalousa}, we uncover a dynamical crossover from isotropic spreading to one-dimensional (1D) constrained growth, which can be tuned via the frustration strength. The kinetic slush state emerges within this crossover regime. Dynamically, it is characterized by dynamical heterogeneity and the real-space coexistence of liquid- and solid-like configurations, while structurally exhibiting fractal geometry alongside generalized Ornstein-Zernike spatial correlations.
Finite-size scaling analysis supports the system's proximity to criticality~\cite{cardy1988FiniteSizeScaling,landau2021Guide}.

\textit{Model}.---
We consider a 2D square lattice of $N=L\times L$ atoms with a lattice constant $a$. Each lattice site hosts a two-level system comprising a ground state $|g\rangle$ and a highly excited Rydberg state $|r\rangle$ [Fig.~\ref{fig1}(a)]. The system is driven by a laser field with Rabi frequency $\Omega$ and detuning $\Delta$. The Hamiltonian of the many-body system reads ($\hbar\equiv1$),
\begin{equation}
		\hat{H} = \frac{\Omega}{2}\sum_j^N\hat{\sigma}_x^j -\Delta\sum_j^N \hat{n}_j + \sum_{j<k}^NV_{jk}\hat{n}_j \hat{n}_k.
\end{equation}
where operator $\hat{\sigma}_j^x=|r_j\rangle\langle g_j|+|g_j\rangle\langle r_j|$ flips the atomic state and $\hat{n}_j=|r_j\rangle\langle r_j|$ is Rydberg density operator of the $j$-th atom.  Atoms in the Rydberg state interact via the vdW potential, ${V}_{jk} = C_6/|\mathbf{R}_j - \mathbf{R}_k|^6$ (where $C_6$ is the state-dependent dispersion coefficient), which describes the two-body interaction between Rydberg atoms at sites $\mathbf{R}_j$ and $\mathbf{R}_k$. Here, $V_{\mathrm{NN}} = C_6/a^6$ denotes the NN interaction strength.

Driven primarily by residual laser phase noise~\cite{lesanovsky2013Kinetic, guardado-sanchez2018Probinga, zhang2024Motionala, Kim_PRX_2024, Hilbert_WB_2026}, this dephasing effect is phenomenologically modeled by a rate $\gamma$ that degrades the coherence between $|g\rangle$ and $|r\rangle$. Because the ground-to-Rydberg coupling is weak, spontaneous decay from the Rydberg state to the ground state is significantly slower than the dephasing process~\cite{lesanovsky2013Kinetic, Hilbert_WB_2026}. Consequently, we focus primarily on dephasing, while spontaneous decay is addressed in the End Matter. In this setting, the dynamics of the dissipative Rydberg system are governed by the Lindblad master equation,
$\dot{\varrho} = \mathcal{L}(\varrho) = -i[\hat{H}, \varrho] + \sum_j \gamma \left( \hat{n}_j \varrho \hat{n}_j - \frac{1}{2} \{ \hat{n}_j, \varrho \} \right)$.

In the strong-dephasing regime ($\gamma\gg\Omega$), adiabatic
elimination of the off-diagonal density-matrix elements yields
a closed evolution equation for the diagonal elements
$\varrho_{\mathbf{n},\mathbf{n}}$~\cite{lesanovsky2013Kinetic,
sibalic2016Drivendissipative,brady2024Anomalousa}.
These elements describe the populations of many-body
configurations $\mathbf{n}=(n_1,\ldots,n_N)$, where $n_j=0$
and $1$ denote the ground and Rydberg states at site $j$,
respectively. Introducing the flipped configuration
$\mathbf{n}^{(j)}=(n_1,\ldots,1-n_j,\ldots,n_N)$,
the effective master equation reads
\begin{equation}
\frac{\partial\varrho_{\mathbf{n},\mathbf{n}}}{\partial\tau}
=\sum_j\Gamma_j(\mathbf{n})
\left[
\varrho_{\mathbf{n}^{(j)},\mathbf{n}^{(j)}}
-\varrho_{\mathbf{n},\mathbf{n}}
\right],
\label{eq1}
\end{equation}
where $\tau=(\Omega^2/\gamma)t$ is the dimensionless time and
the configuration-dependent flip rate is
\begin{equation}\label{rate}
\Gamma_j(\mathbf{n})
=\frac{1}{
1+\frac{4}{\gamma^2}
\left(\Delta-\sum_{k\neq j}V_{jk}n_k\right)^2
}.
\end{equation}
Because $\Gamma_j$ depends only on the occupations at other sites,
flipping $n_j$ leaves the rate unchanged:
$\Gamma_j(\mathbf{n}^{(j)})=\Gamma_j(\mathbf{n})$.
Thus, the laser-driven excitation and de-excitation rates at site $j$
are equal~\cite{lesanovsky2013Kinetic,sibalic2016Drivendissipative}.

To elucidate this facilitation mechanism, we set the laser detuning to compensate for the NN interaction ($\Delta = V_{\mathrm{NN}}$)~\cite{carollo2019Critical,zhang2024Quantum}. Under this setting, the  antiblockade condition is satisfied exclusively when exactly one nearest neighbor is excited ($q_j(\mathbf{n})=\sum_{l\in\mathrm{NN}(j)}n_l=1$). Any deviation ($q_j \neq 1$) introduces a large energetic mismatch of at least $V_{\mathrm{NN}}$, suppressing the transition rate by a factor of $\mathcal{O}(\gamma^2/V_{\mathrm{NN}}^2) \ll 1$ in the interaction-dominated regime. This establishes a strict kinetic constraint characteristic of facilitated epidemic spreading and glass-like growth models~\cite{carollo2019Critical,zhang2024Quantum,perez-espigares2017Epidemic,wintermantel2021Epidemic}.

In a 2D square lattice, the power-law tail of the vdW interaction yields a non-negligible contribution; specifically, the NNN coupling at distance $\sqrt{2}a$ is $V_{\mathrm{NNN}} = V_{\mathrm{NN}}/8$. Within the facilitated sector ($q_j=1$), any excited NNN atom imparts an additional energy shift that detunes the transition from the single-excitation resonance. Truncating the interactions at the NNN level and projecting out the non-facilitated sectors ($q_j \neq 1$), the effective transition rate at site $j$ reduces to
\begin{equation}
\Gamma_j(\mathbf{n})=\delta_{q_j,1}\frac{1}{1+[\eta s_j(\mathbf{n})]^2},
\label{eq:kcm-rate}
\end{equation}
where $s_j(\mathbf{n})=\sum_{m\in\mathrm{NNN}(j)}n_m$ counts the number of excited NNNs, and $\eta \equiv 2V_{\mathrm{NNN}}/\gamma$ parameterizes the kinetic frustration strength. Equation~\eqref{eq:kcm-rate} thus reveals a dual kinetic character: a hard constraint ($\delta_{q_j,1}$) requiring exactly one excited NN to enable transitions, combined with a soft kinetic frustration that progressively slows the local rates as $s_j$ grows. Consequently, the dissipative Rydberg system maps onto a generalized kinetically constrained model~\cite{lesanovsky2013Kinetic,ritort2003Glassy}, where the NNN tail continuously tunes the relaxation timescale without severing the connectivity of the configuration space.  Interactions beyond NNN terms introduce only subleading residual shifts that quantitatively refine the frustration without altering this physical picture [see Supplemental Material (SM)~\cite{SM}].

\textit{Relaxation dynamics}.---To unveil the nonequilibrium nature of the kinetic  state, we investigate its real-time relaxation dynamics. Under strong dephasing, the coherent evolution reduces to a Markovian jump process governed by the constraint-dependent transition rates in Eq.~\eqref{eq:kcm-rate}. We sample the many-body master equation~\eqref{eq1} using a KMC algorithm, with rates from Eq.~\eqref{eq:kcm-rate} evaluated for each instantaneous configuration and updated after every spin flip~\cite{ruizbarlett2009Differences,lesanovsky2013Kinetic,sibalic2016Drivendissipative,brady2024Anomalousa}.
By mapping microscopic transitions to stochastic continuous-time events, this approach efficiently accesses the long relaxation timescales of kinetically constrained systems, enabling direct observation of glassy relaxation and spatiotemporal fluctuations across the lattice.

\begin{figure}
\includegraphics[width=\columnwidth]{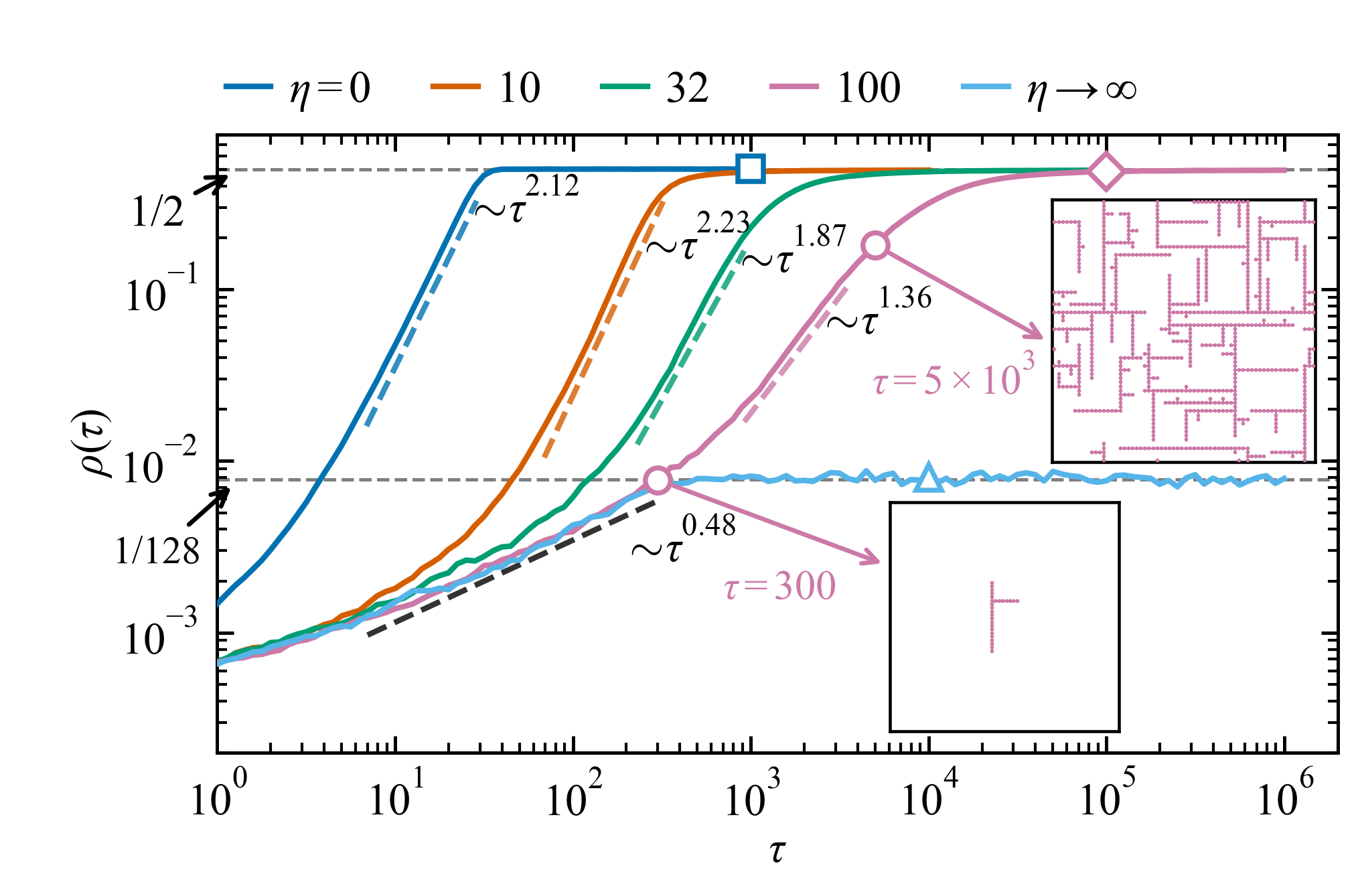}
\caption{\label{fig2} \textbf{Dynamical relaxation under kinetic frustration.} Time evolution of the Rydberg density $\rho(\tau)$ across various frustration strengths $\eta$ on a $64\times64$ lattice. Solid curves denote ensemble averages over 128 independent trajectories, while dashed lines show local power-law fits within the indicated time intervals. Insets display real-space snapshots of a single trajectory at $\tau=300$ and $\tau=5000$ for $\eta=100$. }
\end{figure}

We characterize the excitation dynamics through the mean Rydberg density $\rho(\tau) = L^{-2} \sum_j \langle n_j(\tau) \rangle$, with the brackets denoting an average over independent stochastic trajectories. Initiated from a single Rydberg seed at the lattice center, the evolution exhibits contrasting dynamical regimes governed by the frustration parameter $\eta$ [Fig.~\ref{fig2}]. In the unfrustrated limit ($\eta=0$), $\rho(\tau)$ increases rapidly and quickly reaches a dense, space-filling steady state. A power-law fit over the indicated initial-growth interval yields an effective growth exponent of $2.12$ (blue curve).

At finite frustration, NNN interactions suppress lateral branching at low densities, while facilitated extension along straight chains remains unhindered. Growth therefore proceeds predominantly along linear chains before branching becomes appreciable, producing two growth stages in $\rho(\tau)$. For $\eta=100$ (pink curve), a power-law fit within the early-time regime ($\tau\lesssim300$) yields an effective growth exponent of $0.48$, with similar early-time behavior observed at other large finite values of $\eta$. The corresponding snapshot shows isolated trunk-like filaments [bottom inset, $\tau=300$]. At later times, rare lateral branching events allow the trunks to develop into ramified structures extending across the 2D lattice [top inset, $\tau=5\times10^3$]. The density then grows more rapidly, with an effective growth exponent of $1.36$ over the indicated fitting interval. This intermittent, two-stage evolution reflects the separation of timescales between chain extension and lateral branching (see SM for real-time movies). In the long-time limit, the system ultimately reaches a steady-state density of $\rho_{\mathrm{ss}} = 1/2$, forming a fully percolating network that spans the array.

In the infinitely strong frustration limit ($\eta \to \infty$, light-blue curve), lateral branching is strictly forbidden because facilitation strictly forbids any NNN excitation. This enforces a dimensional reduction from 2D spreading to effective 1D XOR-Fredrickson-Andersen dynamics~\cite{ritort2003Glassy,causer2020}. 
Under this mapping, the domain boundaries execute an unbiased random walk along the chain, leading to the analytical prediction $\rho(\tau)\sim\tau^{1/2}$ (see SM for the derivation). The early-time growth exponent of $0.48$ extracted numerically is consistent with this diffusive scaling. Confined to this 1D manifold, the maximum excitation number is bounded by $L/2$, leading to a size-dependent steady-state density $\rho_{\mathrm{ss}}=1/(2L)$ that vanishes in the thermodynamic limit (e.g., $\rho_{\mathrm{ss}}=1/128$ for $L=64$, indicated by the horizontal dashed line). Movies S1--S3 in the SM illustrate the real-time evolution of excitation patterns in the three regimes: unfrustrated spreading ($\eta=0$), two-stage growth ($\eta=100$), and constrained 1D growth ($\eta\to\infty$).

\begin{figure}[b]
\includegraphics[width=\columnwidth]{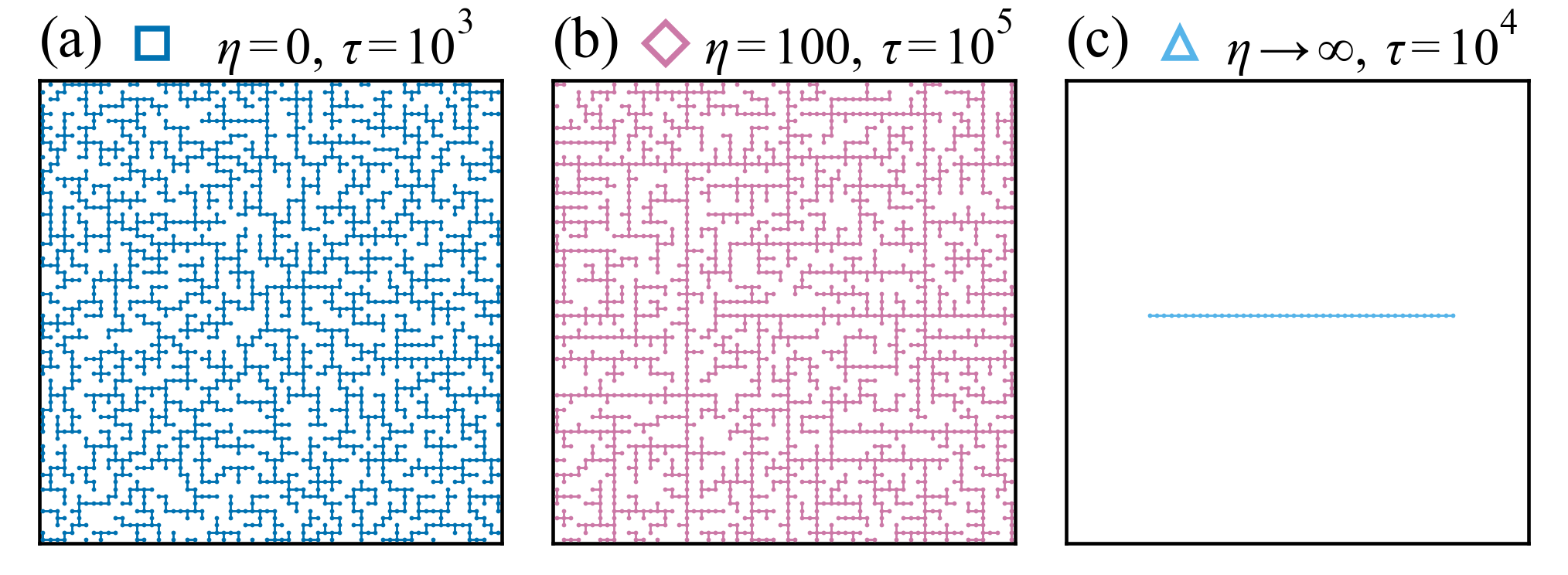}
\caption{\label{fig3} \textbf{Spatial structure of the macroscopic steady state.} (a)--(c) Steady-state spatial configurations on a $64\times64$ lattice for (a) $\eta=0$ at $\tau=10^3$, (b) $\eta=100$ at $\tau=10^5$, and (c) $\eta\to\infty$ at $\tau=10^4$. The symbols in the panel titles correspond to the sampling markers in Fig.~\ref{fig2}. The dots denote Rydberg excitations, with solid lines connecting excited nearest neighbors.}
\end{figure}
\textit{Spatial Structure of the Macroscopic Steady State.}---
After the mean density reaches a stationary plateau, the real-space configurations further distinguish the three kinetic regimes. The snapshots in Fig.~\ref{fig3} illustrate how the spatial morphology changes with the frustration strength.
In the unfrustrated limit ($\eta=0$ at $\tau=10^3$), excitations form a branching network distributed throughout the lattice, with relatively short segments and frequent changes in direction [Fig.~\ref{fig3}(a)].

At finite frustration ($\eta=100$), suppressed lateral branching favors longer, straight chain segments. The configuration at $\tau=10^5$ displays extended trunks connected by branches [Fig.~\ref{fig3}(b)], consistent with the quasi-1D growth followed by delayed branching described above. The resulting system-spanning cluster has a loopless, dendritic geometry, resembling the branching structures reported for quantum slush in closed systems~\cite{zhang2024Quantum}. Related ramified morphologies occur in nonequilibrium growth processes~\cite{vicsek1992fractal}, including diffusion-limited aggregation~\cite{witten1981diffusion}, dendritic crystallization~\cite{langer1980instabilities}, and branching biological networks~\cite{west1997general}.

In the infinite-frustration limit ($\eta\to\infty$), lateral branching is forbidden, and the cluster remains confined to an unbranched 1D chain [Fig.~\ref{fig3}(c), $\tau=10^4$]. Its endpoints continue to fluctuate, allowing the chain to extend and retract. These configurations characterize the nonequilibrium morphologies generated by seeded growth; they do not establish distinct thermodynamic equilibrium phases.
\begin{figure}
\includegraphics[width = \columnwidth]{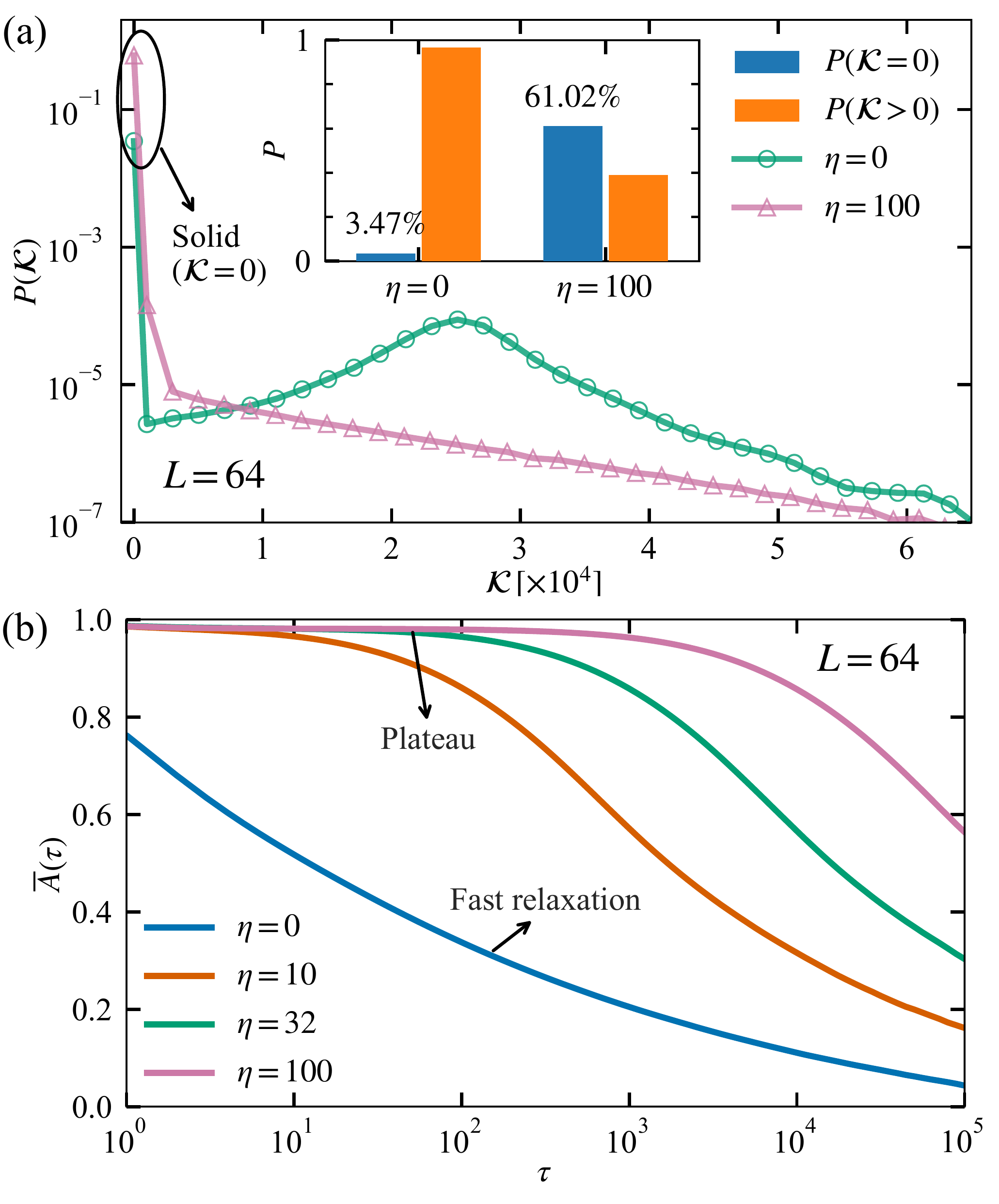}
\caption{\label{fig4} \textbf{Spatio-temporal signatures of the kinetic slush state.} 
(a) Local activity distribution $P(\mathcal{K})$ at $\eta = 0$ and $\eta = 100$. Inset: Comparison between the fractions of sites with no observed flips ($\mathcal{K} = 0$) and sites with observed flips ($\mathcal{K}>0$) within the observation window; the fraction of sites with no observed flips increases from $3.47\%$ to $61.02\%$ as frustration is introduced. Statistics are accumulated over 256 independent KMC trajectories within an observation window of $T=10^5$. 
(b) Site-averaged autocorrelation function $\bar{A}(\tau)$ for various $\eta$, showing slower relaxation with increasing $\eta$ and an extended plateau followed by slow decay at strong frustration.}
\end{figure}

\textit{Dynamical heterogeneity and kinetic slush.}---
Even after the global density $\rho(\tau)$ reaches a stationary plateau [Fig.~\ref{fig2}], local spin-flip dynamics remain active. To quantify these microscopic dynamics, we evaluate the site-resolved dynamical activity $\mathcal{K}_j$, defined as the total number of spin flips at site $j$ over an observation window within the density-plateau regime~\cite{lesanovsky2013Kinetic}. The resulting activity map reveals strong spatial heterogeneity: highly active, liquid-like regions (red/yellow) are interwoven with a low-activity, solid-like network (dark blue) [Fig.~\ref{fig1}(c)]. This low-activity network  follows the loopless dendritic structure shown in Fig.~\ref{fig1}(b).

To quantify this spatial contrast, we analyze the local activity distribution $P(\mathcal{K})$ over an observation window of $T=10^5$ [Fig.~\ref{fig4}(a)]. We distinguish sites with no observed flips ($\mathcal{K}=0$) from those with at least one flip ($\mathcal{K}>0$). The fraction of sites with no observed flips increases from $3.47\%$ at $\eta=0$ to $61.02\%$ at $\eta=100$ [inset, Fig.~\ref{fig4}(a)]. For $\mathcal{K}>0$, the distribution at $\eta=0$ exhibits a peak at finite activity. At $\eta=100$, a large probability mass at $\mathcal{K}=0$ coexists with a broad nonzero-activity distribution extending to large $\mathcal{K}$. Thus, some sites remain highly active alongside sites with no observed flips, providing a dynamical signature of the kinetic slush state.

Complementing the activity distribution, we probe temporal relaxation through the site-averaged autocorrelation function $\bar{A}(\tau)=L^{-2}\sum_j\langle\sigma_j^z(\tau_0)\sigma_j^z(\tau_0+\tau)\rangle$, where $\sigma_j^z=2n_j-1$ is the local spin and $\langle\dots\rangle$ denotes an average over time origins $\tau_0$ within the density-plateau regime. The unfrustrated system ($\eta=0$) relaxes rapidly, whereas increasing $\eta$ substantially delays the decay of $\bar{A}(\tau)$ [Fig.~\ref{fig4}(b)]. At strong frustration, the autocorrelation exhibits an extended plateau followed by a slow decay, indicating that local spin states remain correlated with their initial values over long times. This slow relaxation, together with the heterogeneous activity distribution $P(\mathcal{K})$, supports glass-like dynamics in the kinetic slush state~\cite{rau2016Spin,gao2025Unified}.

\begin{figure}[h!]
\includegraphics[width=\columnwidth]{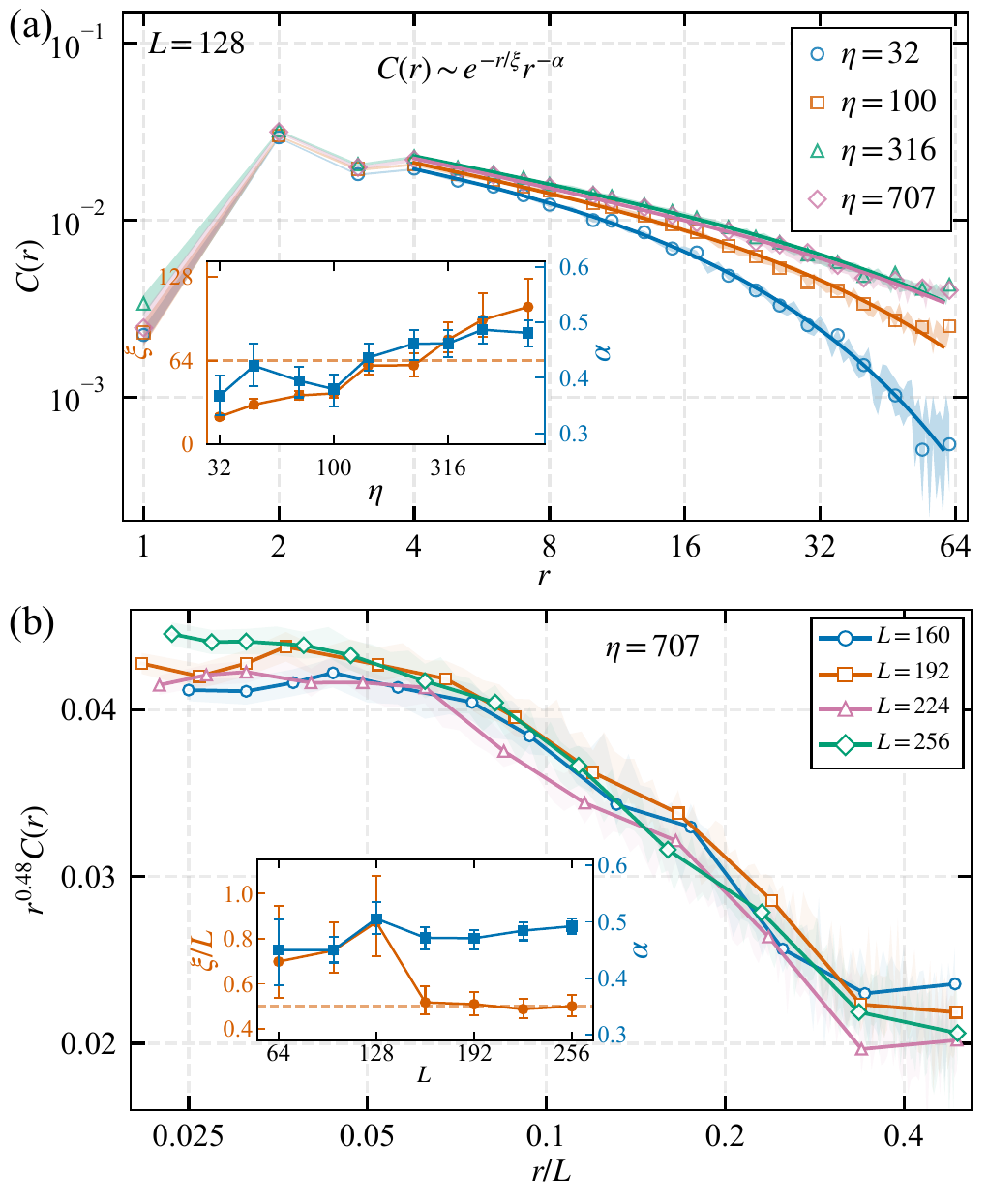}
\caption{\label{fig5} 
	\textbf{Spatial correlations and finite-size scaling analysis.}
	(a) Spatial correlation function $C(r)$ for $L=128$ at different frustration strengths $\eta$. Solid curves show fits to the generalized Ornstein-Zernike form. Inset: Fitted correlation length $\xi$ (red) and exponent $\alpha$ (blue) versus $\eta$; the dashed line marks $\xi=L/2$.
	(b) Finite-size scaling of $C(r)$ at $\eta=707$. The rescaled correlations $r^{0.48}C(r)$, plotted against $r/L$ for different system sizes $L$, collapse over an intermediate range of scaled distances. Inset: Normalized correlation length $\xi/L$ (red) and exponent $\alpha$ (blue) versus $L$; the dashed line marks $\xi/L=0.5$. Shaded bands and error bars indicate standard errors estimated via bootstrapping over 128 independent trajectories.}
\end{figure}

\textit{Emergent criticality and spatial correlations.}---
We next quantify the spatial organization of kinetic slush through the equal-time density correlation function,
$C(\mathbf{r}) = L^{-2}\sum_j[\langle n_j n_{j+\mathbf{r}}\rangle-\langle n_j\rangle\langle n_{j+\mathbf{r}}\rangle]$.
We analyze its decay along the lattice $x$ axis, defining $C(r)\equiv C(r,0)$.
As shown in Fig.~\ref{fig5}(a), the long-distance decay of $C(r)$ becomes slower as the frustration parameter $\eta$ increases. To characterize this decay, we fit the numerical data to the generalized Ornstein-Zernike form $C(r)\propto r^{-\alpha}e^{-r/\xi}$, where $\xi$ is the fitted correlation length and $\alpha$ governs the algebraic prefactor~\cite{Hinrichsen01112000}. The inset of Fig.~\ref{fig5}(a) shows that $\xi$ grows with increasing frustration, while $\alpha$ approaches $0.48$ in the strong-frustration regime. For $\eta\gtrsim316$, the fitted correlation length exceeds $L/2$, the maximum separation along a periodic lattice axis, indicating system-scale spatial correlations.
In this regime, we determine the fractal dimension by scaling the cluster mass with chemical distance, obtaining $d_f=1.808\pm0.005$ (see SM). This value is close to $d_f\approx1.798$ reported for quantum slush in closed systems~\cite{zhang2024Quantum}. The similarity suggests that kinetic constraints play an important role in shaping the branching geometry in both coherent and dissipative systems.

To examine the finite-size scaling of these system-scale correlations, we vary the system size $L$ at a fixed frustration strength of $\eta=707$ [Fig.~\ref{fig5}(b)]. The inset shows that $\alpha$ approaches $0.48$ for larger systems. While the fitted correlation length exceeds $L/2$ for smaller systems, the ratio $\xi/L$ approaches $0.5$ for $L\gtrsim160$, indicating approximately linear scaling of $\xi$ with $L$ over this size range.
With $\alpha\simeq0.48$ and $\xi\simeq L/2$, the fitted correlation function takes the approximate form $C(r,L)\propto r^{-0.48}\exp(-2r/L)$. Accordingly, plotting $r^{0.48}C(r,L)$ against $r/L$ yields a good data collapse for $L=160,192,224,$ and $256$ over an intermediate range of scaled distances [Fig.~\ref{fig5}(b)]. This collapse is consistent with the finite-size scaling form $C(r,L)=r^{-0.48}\mathcal{F}(r/L)$~\cite{cardy1988FiniteSizeScaling,Hinrichsen01112000,landau2021Guide}. Together, the approximately constant ratio $\xi/L$ and the correlation-function collapse indicate system-scale correlations and support proximity to criticality for the system sizes studied.

\textit{Discussion and Conclusion.}---
In summary, we have investigated the formation and dynamics of kinetic slush in dissipative Rydberg arrays. The underlying mechanism is a competition between facilitated chain extension and NNN-induced suppression of lateral branching. The resulting separation of timescales between quasi-1D growth and delayed 2D branching produces ramified clusters. Within a finite observation window, the local activity distribution reveals the coexistence of sites with frequent flips and sites with no observed flips, while the autocorrelation exhibits an extended plateau followed by slow decay. These observations support dynamically heterogeneous, glass-like relaxation. The cluster geometry and spatial correlations further characterize the state: the  fractal dimension is close to that reported for quantum slush, and the correlation functions exhibit finite-size scaling consistent with proximity to criticality.

\begin{figure*}[t]
\centering
\includegraphics[width=0.68\textwidth]{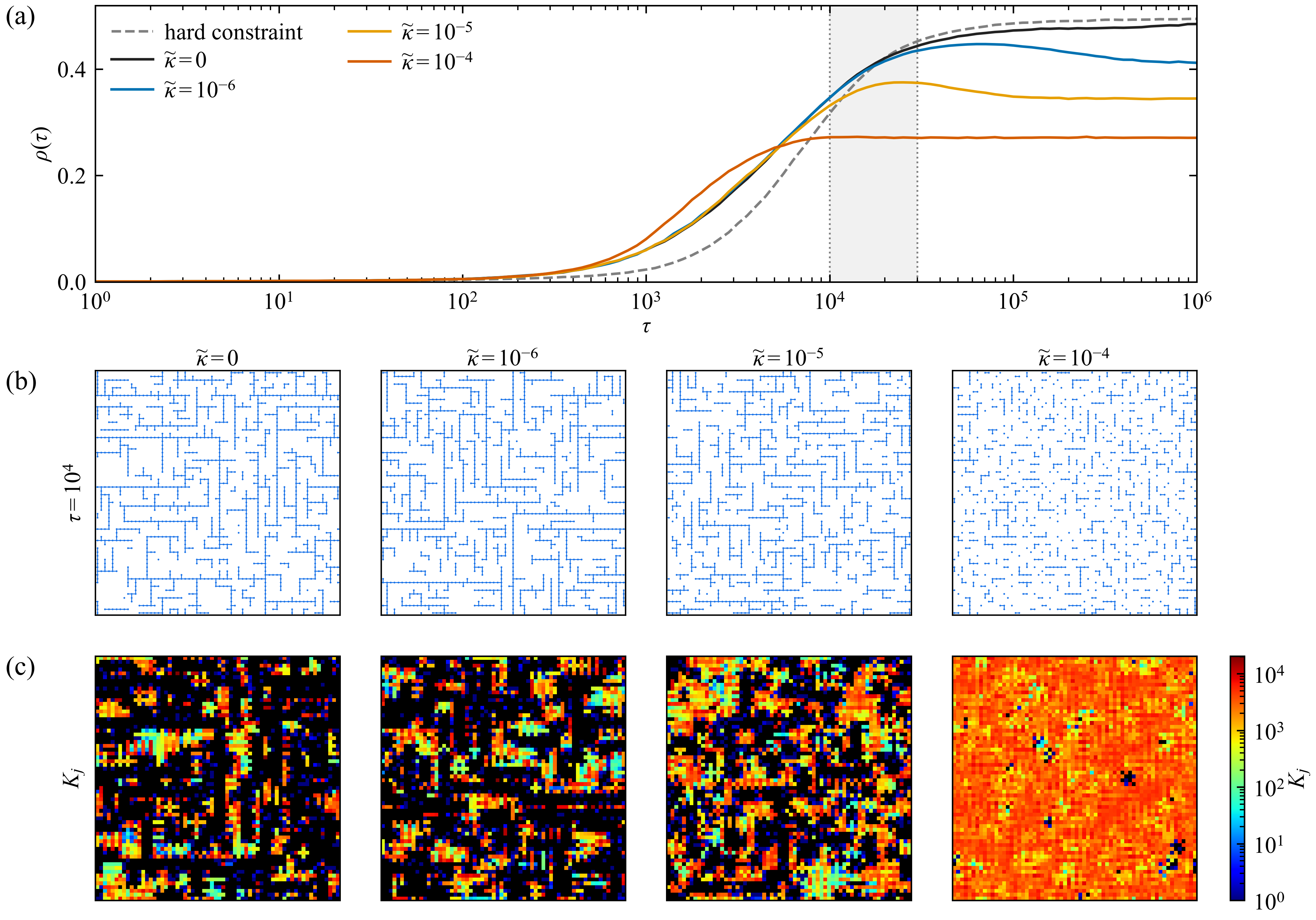}
\caption{\label{fig:realistic-dynamics}
\textbf{Kinetic slush dynamics beyond the hard-constraint limit.}
(a) Time evolution of the mean Rydberg density $\rho(\tau)$ at rescaled decay rates $\tilde{\kappa}=0,10^{-6},10^{-5},$ and $10^{-4}$ (solid lines), compared with the hard-constraint benchmark (dashed gray line). The shaded interval, $10^4\leq\tau\leq3\times10^4$, marks the observation window used in (c).
(b) Excitation configurations at $\tau=10^4$, showing increasing fragmentation of the ramified clusters as $\tilde{\kappa}$ increases.
(c) Local dynamical activity $K_j$ accumulated over the shaded interval in (a), showing reduced spatial activity contrast at stronger decay.
Simulation parameters: $L=64$ and $\eta=100$. Density curves in (a) are averaged over 128 trajectories.}
\end{figure*}

The loopless, dendritic structures resemble ramified patterns found in nonequilibrium growth processes such as diffusion-limited aggregation and dendritic crystallization~\cite{vicsek1992fractal,Fractal_Mandelbrot}. Their emergence in the strong-dephasing regime suggests that kinetic constraints can generate slush-like structural and dynamical features even when coherent quantum evolution is suppressed.
Programmable neutral-atom optical tweezer arrays provide a setting for testing these predictions~\cite{browaeys2020Manybody,bluvstein2022quantum,chen2023Continuous,manetsch2025tweezer,chen2025Collective,zhang2025observation}. 
As detailed in the End Matter [Fig.~\ref{fig:realistic-dynamics}], ramified structures and dynamical heterogeneity persist within the observation window when off-resonant transitions and weak spontaneous decay are included. Stronger decay fragments the clusters and reduces the activity contrast. Experimental observation therefore requires a formation and measurement window short compared with the decay timescale.

\textit{Acknowledgements.}---
This work was supported by the National Natural Science Foundation of China (12347102, 12675041, 12274131,  and 12547183), the Natural Science Foundation of Jiangsu Province (BK20233001), the Innovation Program for Quantum Science and Technology (2024ZD0300101), and the Fundamental and Interdisciplinary Disciplines Breakthrough Plan of the Ministry of Education of China (JYB2025XDXM502).

%


\onecolumngrid

\vspace{0.5em}
\begin{center}
{\bfseries End Matter}
\end{center}
\vspace{0.5em}

\twocolumngrid

\textit{The physical realization}.---
We include spontaneous decay of Rydberg excitations at rate $\kappa$—an unavoidable dissipative channel that becomes relevant over long timescales $\sim\kappa^{-1}$. Decay from the Rydberg state to the ground state is modeled by adding the local Lindblad dissipator
$\mathcal{D}[\varrho]=\kappa\sum_j\left(\sigma_j^-\varrho\sigma_j^+-\frac{1}{2}\{\hat{n}_j,\varrho\}\right)$
to the master equation, where $\sigma_j^-=|g_j\rangle\langle r_j|$.
We also relax the strict single-NN facilitation constraint in Eq.~\eqref{eq:kcm-rate} by restoring the full Lorentzian transition rate while retaining NN and NNN interactions. At the facilitation resonance $\Delta=V_{\mathrm{NN}}$, using $V_{\mathrm{NN}}=8V_{\mathrm{NNN}}$, the dimensionless laser-induced transition rate becomes
\begin{equation}
	\Gamma_j(\mathbf{n})=\frac{1}{1+\eta^2[8(1-q_j)-s_j]^2}.
	\label{eq:full-realistic-rate}
\end{equation}
For $q_j=1$, this expression recovers the facilitated rate in Eq.~\eqref{eq:kcm-rate}. For $q_j\neq1$, it retains the residual off-resonant transitions excluded by the hard constraint.

Spontaneous decay at rate $\kappa$ explicitly breaks detailed balance. In the strong-dephasing regime ($\gamma\gg\Omega,\kappa$), adiabatic elimination of the quantum coherences yields the effective classical jump rates
\begin{equation}
\begin{aligned}
w_j^{0\to1}(\mathbf{n})&=\Gamma_j(\mathbf{n}),\\
w_j^{1\to0}(\mathbf{n})&=\Gamma_j(\mathbf{n})+\tilde\kappa,
\end{aligned}
\label{eq:realistic-decay-rates}
\end{equation}
where $\tilde\kappa=\kappa\gamma/\Omega^2$ is the rescaled decay rate. The many-body evolution generalizes Eq.~\eqref{eq1} to include spontaneous downward jumps, with total flip rate $w_j(\mathbf{n})=\Gamma_j(\mathbf{n})+\tilde\kappa n_j$. The laser induces excitation and de-excitation at equal rates $\Gamma_j$, whereas spontaneous decay adds a downward transition independent of the local facilitation condition. The associated decay timescale, $\tilde\kappa^{-1}$, sets the scale over which radiative losses compete with constrained growth.

Off-resonant laser transitions and spontaneous decay modify the cluster connectivity through different mechanisms. Rare off-resonant excitations can create isolated seeds without an excited nearest neighbor, while transitions at sites with multiple excited nearest neighbors can connect existing branches, enabling cluster merging and loop formation. Spontaneous decay can instead remove excitations that connect different parts of a cluster, breaking these connections and fragmenting the ramified structure.

We first compare the full-rate dynamics with the hard-constraint benchmark, starting from a single seed on a $64\times64$ periodic lattice at $\eta=100$ [Fig.~\ref{fig:realistic-dynamics}(a)]. Without spontaneous decay ($\tilde\kappa=0$), the density evolution retains the two-stage growth pattern of the hard-constrained model, with slow quasi-1D chain extension followed by faster growth associated with 2D branching. The density begins to increase earlier than in the hard-constraint benchmark, consistent with additional seeds nucleating through rare off-resonant excitations. Spontaneous decay modifies the subsequent evolution: for $\tilde\kappa=10^{-6}$ and $10^{-5}$, the density reaches a maximum before gradually decreasing, whereas at $\tilde\kappa=10^{-4}$ it approaches a lower plateau. Overall, increasing the decay rate lowers the late-time Rydberg density.

At $\tau=10^4$, the configurations at weak decay retain extended chains and ramified branches [Fig.~\ref{fig:realistic-dynamics}(b)]. We quantify the spatial variation in the dynamics through the local activity $K_j$, accumulated over the observation window $10^4\leq\tau\leq3\times10^4$ highlighted in Fig.~\ref{fig:realistic-dynamics}(a). For $\tilde\kappa\leq10^{-5}$, low-activity regions coexist with localized regions of frequent spin flips [Fig.~\ref{fig:realistic-dynamics}(c)]. Thus, the structural and dynamical signatures of kinetic slush persist within the observation window despite off-resonant transitions and weak spontaneous decay.

At $\tilde\kappa=10^{-4}$, the observation times become comparable to the decay timescale $\tilde\kappa^{-1}$. The configuration at $\tau=10^4$ consists predominantly of disconnected fragments [rightmost panel, Fig.~\ref{fig:realistic-dynamics}(b)], consistent with spontaneous decay removing excitations that connect different parts of the cluster. The corresponding activity map shows a broadly active background with fewer low-activity regions and reduced spatial contrast [Fig.~\ref{fig:realistic-dynamics}(c)]. For the parameters considered, these results indicate that kinetic slush features survive over a finite time window at weak decay, while cluster connectivity and dynamical heterogeneity are reduced when the evolution time becomes comparable to the decay timescale.

Neutral-atom platforms provide a direct setting to experimentally
explore the predicted kinetic slush dynamics. As an illustrative
parameter set for rubidium arrays, we consider two-photon
excitation to $nS$ Rydberg states via the intermediate
$6P_{3/2}$ state, with $\Omega\simeq5\,\mathrm{MHz}$ and
$\gamma\simeq25\,\mathrm{MHz}$.
Accessing frustration strengths $\eta=10\text{--}100$ requires
NN interactions
$V_{\mathrm{NN}}=C_6/a^6\simeq1\text{--}10\,\mathrm{GHz}$.
At lattice spacings of $a\simeq3\text{--}4\,\mu\mathrm{m}$,
the approximate scaling $C_6\propto n^{11}$ motivates considering
high-$n$ Rydberg states~\cite{sibalic2017arc}.
The rescaled decay rates of interest correspond to physical
decay rates $\kappa\simeq0.1\text{--}1\,\mathrm{kHz}$.
High-$n$ Rydberg levels or long-lived circular states may
be considered to reduce spontaneous decay~\cite{
	Long-lived_Cantat_2020,Interacting_Circular_2025}.
The corresponding characteristic decay times are
$t_{\mathrm{decay}}=\kappa^{-1}\simeq1\text{--}10\,\mathrm{ms}$.
This suggests millisecond-scale observation windows for probing the heterogeneous dynamics.

\clearpage
\onecolumngrid
\setcounter{equation}{0}
\setcounter{figure}{0}
\setcounter{table}{0}
\setcounter{section}{0}
\setcounter{secnumdepth}{2}
\renewcommand{\theequation}{S\arabic{equation}}
\renewcommand{\thefigure}{S\arabic{figure}}
\renewcommand{\thetable}{S\Roman{table}}
\renewcommand{\theHequation}{supp.\arabic{equation}}
\renewcommand{\theHfigure}{supp.\arabic{figure}}
\renewcommand{\theHtable}{supp.\arabic{table}}
\renewcommand{\theHsection}{supp.\arabic{section}}
\begin{center}
{\large\bfseries Supplemental Material for\\[2mm]
Kinetic Slush in Dissipative Rydberg Arrays}
\par\vspace{3mm}
Hufeng Wen$^{1}$, Ya-Xin Xiang$^{1}$, Qun-Li Lei$^{1,2,4}$,
Thomas Pohl$^{3}$, Zhengyang Bai$^{1,2}$, and Yu-Qiang Ma$^{1,2,4}$
\par\vspace{2mm}
{\small\itshape
$^1$National Laboratory of Solid State Microstructures and School of Physics, Collaborative Innovation Center of Advanced Microstructures, Nanjing University, Nanjing 210093, China\\
$^2$Jiangsu Physical Science Research Center\\
$^3$Institute for Theoretical Physics, TU Wien, Vienna 1040, Austria\\
$^4$Hefei National Laboratory, Hefei 230088, China}
\end{center}

In this Supplemental Material, we derive the effective transition rates for Rydberg arrays in the strong-dephasing regime and examine the effects of van der Waals interactions  beyond next-nearest
neighbors on the slush dynamics. We also analyze the early-time density growth in the infinite-frustration limit. Finally, we characterize the spatial correlations and cluster geometry using the static structure factor and the fractal dimension.

\section{Derivation of Effective Rate Equations}
\label{sec:rate-equation}
We derive the configuration-dependent flip rates used in our kinetic Monte Carlo (KMC) simulations for a lattice of $N$ driven two-level atoms, with ground state $|g_j\rangle$ and Rydberg state $|r_j\rangle$. Setting $\hbar=1$, the Hamiltonian is
\[
\hat H=\frac{\Omega}{2}\sum_j\hat\sigma_j^x
-\Delta\sum_j\hat n_j+\sum_{j<k}V_{jk}\hat n_j\hat n_k,
\]
where $\hat n_j=|r_j\rangle\langle r_j|$, $\hat\sigma_j^x=|r_j\rangle\langle g_j|+|g_j\rangle\langle r_j|$, $\Omega$ is the Rabi frequency, and $\Delta$ is the laser detuning. The interaction is $V_{jk}=C_6/|\mathbf R_j-\mathbf R_k|^6$, where $\mathbf R_j$ is the position of atom $j$. In this derivation, we neglect spontaneous decay and retain dephasing at rate $\gamma$. The many-body density matrix $\varrho$ evolves according to
\begin{equation}\label{eq:master}
\dot{\varrho} = -i[\hat{H}, \varrho] + \gamma \sum_j \left( \hat{n}_j \varrho \hat{n}_j - \frac{1}{2} \{ \hat{n}_j, \varrho \} \right).
\end{equation}
In the strong-dephasing regime ($\gamma\gg\Omega$), the coherence relaxation time, of order $\gamma^{-1}$, is much shorter than the characteristic local population-transfer time, of order $\gamma/\Omega^2$. Adiabatic elimination of the off-diagonal density-matrix elements therefore yields a closed evolution equation for the diagonal elements $\varrho_{\mathbf{n},\mathbf{n}}$~\cite{lesanovsky2013Kinetic,sibalic2016Drivendissipative,brady2024Anomalousa}. These elements describe the populations of many-body configurations $\mathbf{n}=(n_1,\ldots,n_N)$, where $n_j=0$ and $1$ denote the ground and Rydberg states at site $j$, respectively. Introducing the flipped configuration $\mathbf{n}^{(j)}=(n_1,\ldots,1-n_j,\ldots,n_N)$, the effective master equation reads
\begin{equation}\label{eq:rate_eq}
\frac{\partial\varrho_{\mathbf{n},\mathbf{n}}}{\partial\tau}
=\sum_j\Gamma_j(\mathbf{n})
\left[\varrho_{\mathbf{n}^{(j)},\mathbf{n}^{(j)}}-\varrho_{\mathbf{n},\mathbf{n}}\right],
\end{equation}
where $\tau=(\Omega^2/\gamma)t$ is the dimensionless time and the configuration-dependent flip rate is
\begin{equation}\label{eq:gamma_j}
\Gamma_j(\mathbf{n})=
\frac{1}{1+\frac{4}{\gamma^2}\left(\Delta-\sum_{k\neq j}V_{jk}n_k\right)^2}.
\end{equation}
Because $\Gamma_j$ depends only on the occupations at other sites, flipping $n_j$ leaves the rate unchanged: $\Gamma_j(\mathbf{n}^{(j)})=\Gamma_j(\mathbf{n})$. Thus, the laser-driven excitation and de-excitation rates at site $j$ are equal. For the following interaction truncation, we define $\tilde\Delta=2\Delta/\gamma$ and $\tilde V_{jk}=2V_{jk}/\gamma$.

The KMC algorithm samples Eq.~\eqref{eq:rate_eq} by stochastic flips of the binary occupations. At each event, the transition at site $j$ has rate
\begin{equation}\label{eq:KMC_rate}
W_j(\mathbf{n})\equiv
W(n_j\to1-n_j\mid\mathbf{n}_{\setminus j})=\Gamma_j(\mathbf{n}),
\end{equation}
where $\mathbf{n}_{\setminus j}$ denotes the occupations at all other sites. The rates are evaluated using the instantaneous microscopic configuration and updated after each event.

For the two-dimensional square-lattice model considered here, we expand the interaction-induced level shift in contributions from successive neighbor shells,
\begin{equation}
\sum_{k\neq j}\tilde V_{jk}n_k
=\tilde V_{\mathrm{NN}}q_j
+\tilde V_{\mathrm{NNN}}s_j
+\tilde V_{\mathrm{3NN}}u_j+\cdots,
\label{eq:neighbor-expansion}
\end{equation}
where
\begin{equation}
q_j=\sum_{l\in\mathrm{NN}(j)}n_l,\qquad
s_j=\sum_{m\in\mathrm{NNN}(j)}n_m,\qquad
u_j=\sum_{p\in\mathrm{3NN}(j)}n_p
\end{equation}
count the excited nearest-neighbor (NN), next-nearest-neighbor (NNN), and third-nearest-neighbor (3NN) atoms, respectively. For isotropic van der Waals interactions, the distances $a$, $\sqrt{2}a$, and $2a$ give $V_{\mathrm{NNN}}=V_{\mathrm{NN}}/8$ and $V_{\mathrm{3NN}}=V_{\mathrm{NNN}}/8$.

Setting $\Delta=V_{\mathrm{NN}}$ and defining $\eta=2V_{\mathrm{NNN}}/\gamma$, the configuration-dependent rate becomes
\begin{equation}
\Gamma_j(\mathbf n)
=\frac{1}{1+\eta^2\left[8(1-q_j)-s_j-\frac{u_j}{8}-\cdots\right]^2}.
\label{eq:full-shell-rate}
\end{equation}
In the interaction-dominated regime, transitions with $q_j\neq1$ are far off resonance. Imposing the hard single-NN facilitation constraint used in the main text gives
\begin{equation}
\Gamma_j(\mathbf n)
=\frac{\delta_{q_j,1}}{1+\eta^2\left(s_j+\frac{u_j}{8}+\cdots\right)^2}.
\label{eq:constrained-shell-rate}
\end{equation}
The Kronecker delta restricts flips to sites with exactly one excited NN. In the main text, we retain the NN and NNN contributions to focus on the kinetic frustration induced by NNN interactions.
To test the leading correction to the NNN truncation, we retain the 3NN term in Eq.~\eqref{eq:constrained-shell-rate}. We perform KMC simulations with this rate on a $64\times64$ lattice at $\eta=100$, retaining the hard single-NN constraint.

\begin{figure}[htbp]
\centering
\includegraphics[width=0.75\linewidth]{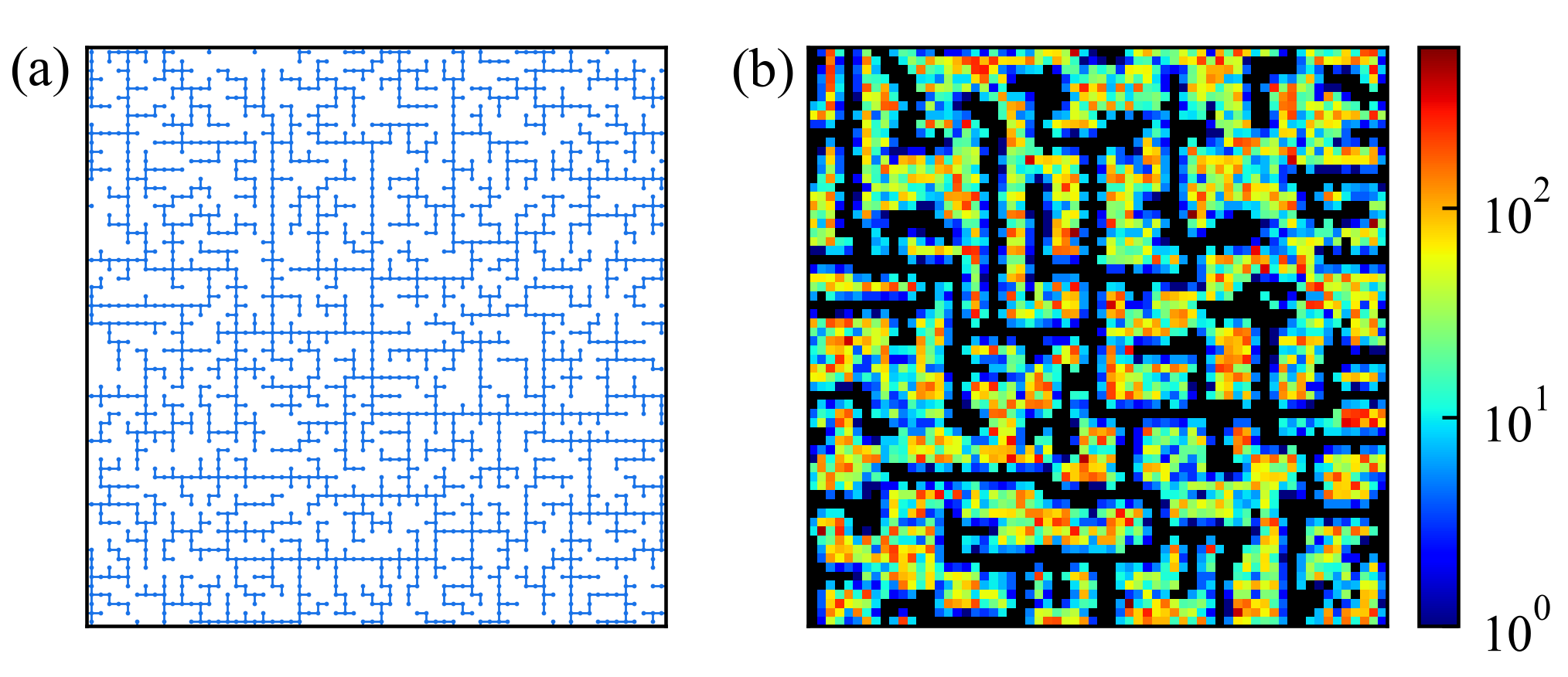}
\caption{\label{figs1}
\textbf{Effect of the 3NN interaction on kinetic slush.} (a) A representative late-time configuration with a loopless, ramified geometry. (b) Corresponding local dynamical activity $\mathcal{K}_j$, defined as the number of spin flips at site $j$ over a late-time observation window of $T=10^6$. Low-activity regions coexist with regions of frequent flips. Simulations retain interactions through 3NN and the hard single-NN constraint, with $L=64$ and $\eta=100$.
}
\end{figure}

The representative configuration in Fig.~\ref{figs1}(a) retains a loopless, ramified geometry. The local activity map in Fig.~\ref{figs1}(b) also shows spatial heterogeneity, with low-activity regions coexisting with regions of frequent spin flips. Thus, including the 3NN interaction preserves the main structural and dynamical features identified in the NNN-truncated model, supporting the conclusions of the main text under the conditions examined here.

\section{Effective One-Dimensional Description of Frustrated Dynamics}

The strict-frustration limit ($\eta\to\infty$) in the main text provides an analytic benchmark for understanding the slow early-time density growth observed in the strongly frustrated regime (e.g., the $\eta = 100$ case shown in Fig.~2 of the main text). In this infinite-$\eta$ limit, a site can flip only if none of its NNN sites is excited. To obtain an effective one-dimensional (1D) benchmark, we restrict the cluster to a fixed lattice axis. The resulting dynamics describe the extension and retraction of an unbranched chain. This characteristic linear geometry is clearly visible in the early-time snapshot of the $\eta=100$ dynamics at $\tau=300$.

{ The allowed flips along an unbranched chain follow the symmetric 1D XOR-Fredrickson--Andersen (XOR-FA) rule~\cite{causer2020}. As derived below, the resulting diffusive motion of the domain boundaries yields a density growth exponent of $1/2$. For large but finite frustration, this description provides a reference for the early-time dynamics before lateral branching leads to two-dimensional growth.}

Consider an $L\times L$ array initialized with a single seed. In the strict-frustration limit, the resulting unbranched trunk can be represented by a contiguous cluster of length $l\in\{1,2,\ldots,L-1\}$ on a periodic 1D chain of length $L$. The one-NN constraint restricts state flips strictly to the two cluster tips: an adjacent empty site can be excited, or an occupied tip can be de-excited. Each allowed tip transition occurs at a unit rate in the rescaled time $\tau$. Consequently, away from the interval endpoints, the aggregate transition rates for both $l\to l+1$ and $l\to l-1$ are exactly $2$. Furthermore, the kinetic constraint enforces $l=1$ and $l=L-1$ as reflecting boundaries, ensuring that the trunk can neither vanish entirely nor fully occupy the system. Letting $P_l(\tau)$ denote the probability of observing a trunk of length $l$, we derive the following probability-balance equations:
\begin{equation}
\begin{aligned}
\frac{dP_l}{d\tau}
&=2P_{l-1}+2P_{l+1}-4P_l,
&&1<l<L-1,\\
\frac{dP_1}{d\tau}
&=2P_2-2P_1,\\
\frac{dP_{L-1}}{d\tau}
&=2P_{L-2}-2P_{L-1} .
\end{aligned}
\label{eq:length-balance}
\end{equation}
Since the trunk occupies $l$ sites in the underlying $L\times L$ array, the  density is given by $\rho(\tau)=\langle l(\tau)\rangle/L^2$, where $\langle l\rangle=\sum_{l=1}^{L-1}lP_l$. Multiplying Eq.~\eqref{eq:length-balance} by $l$ and summing over all $l$ pairwise cancels the unbiased bulk currents, yielding the boundary-current relation:\begin{equation}\frac{d\langle l\rangle}{d\tau}=2P_1(\tau)-2P_{L-1}(\tau).\label{eq:boundary-current}\end{equation}Given the single-seed initial condition, $P_l(0)=\delta_{l,1}$. In the diffusive regime $1\ll\tau\ll L^2$, the diffusive front has not yet reached the upper boundary, meaning $P_{L-1}\simeq0$. The process thus maps onto diffusion on a semi-infinite line with a reflecting boundary, where $P_1(\tau)\simeq(2\pi\tau)^{-1/2}$. Keeping only the leading diffusive contribution in Eq.~\eqref{eq:boundary-current}, we obtain
\begin{equation}
\rho(\tau)-\rho(0)
\sim\frac{2}{L^2}\sqrt{\frac{2\tau}{\pi}},
\qquad 1\ll\tau\ll L^2.
\label{eq:early-density-growth}
\end{equation}
The numerically extracted growth exponent of $0.48$ in the $\eta\to\infty$ limit is consistent with the predicted diffusive exponent of $1/2$ over the fitted time interval.

At times of order $L^2$, the upper reflecting boundary affects the length distribution, and the mean density approaches a finite-size plateau. In this fixed-axis 1D model, the reflected random walk has a uniform stationary length distribution:\begin{equation}P_l(\tau\to\infty)=\frac{1}{L-1},\qquad\langle l\rangle(\tau\to\infty)=\frac{L}{2},\qquad\rho(\tau\to\infty)=\frac{1}{2L}.\label{eq:density-saturation}\end{equation}
Here, $L/2$ is the stationary mean length, whereas the maximum allowed length is $L-1$.
Consequently, in the strict-frustration limit, the total number of excitations scales as $O(L)$ rather than $O(L^2)$, causing the steady-state density to vanish as $L^{-1}$. This $L^{-1}$ scaling is consistent with the size-dependent density plateau discussed in the main text.

\section{Static Spatial Structure of the Kinetic Slush State}
\subsection{Structure Factor of the Kinetic Slush state}

\begin{figure}[h!]
	\includegraphics[width=0.65\textwidth]{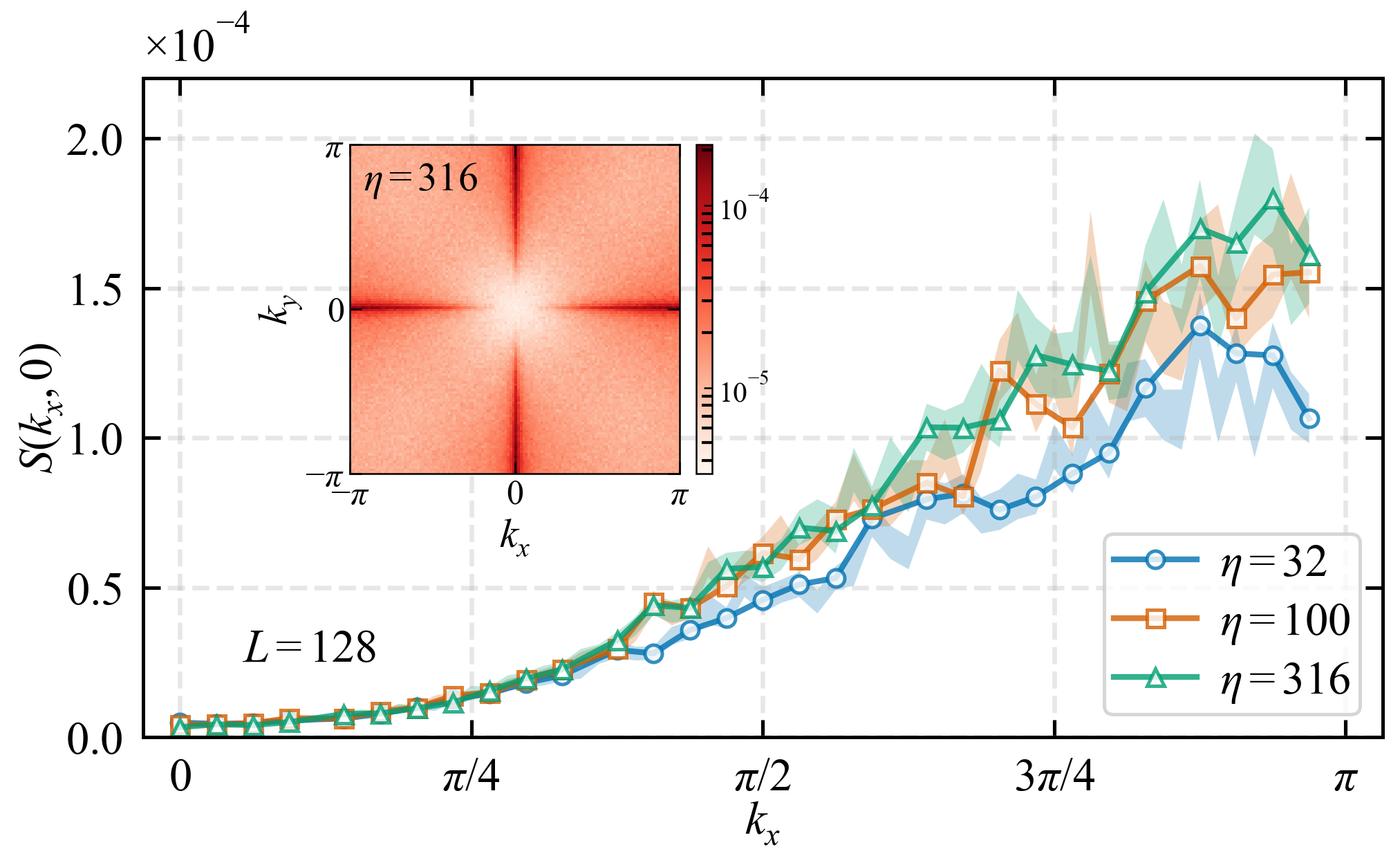}
	\caption{\label{figs2}
\textbf{Static structure factor of the kinetic slush state.}
The main panel shows the slice $S(k_x,0)$ for different frustration strengths $\eta$. Inset: Two-dimensional structure factor $S(\mathbf{k})$
at $\eta=316$, showing a cross-shaped pattern indicative
of anisotropic spatial correlations. Results are averaged over $128$ independent trajectories
on a $128\times128$ lattice. Shaded bands indicate statistical uncertainties.
	}
\end{figure}
To further quantify the spatial correlations $C(\mathbf{r})$ introduced in the main text, we calculate the static structure factor, $S(\mathbf{k}) = \frac{1}{L^2} \sum_{\mathbf{r}} C(\mathbf{r}) e^{i\mathbf{k}\cdot\mathbf{r}}$. As shown in the inset of Fig.~\ref{figs2}, the two-dimensional structure factor displays a highly anisotropic, cross-shaped profile, with enhanced intensity along the $k_x=0$ and $k_y=0$ axes. This anisotropic spectral pattern is consistent with the preferential growth along the lattice axes discussed in the main text.  Analyzing the 1D cross section $S(k_x, 0)$ in the main panel of Fig.~\ref{figs2} reveals a peak near $k_x \approx \pi$. The enhanced structure-factor intensity near the Brillouin-zone boundary indicates a short-wavelength component of the spatial correlations. At the smallest wave vectors, $S(k_x,0)$ takes similar values of order $4\times10^{-6}$ across the frustration strengths considered. This indicates that long-wavelength density fluctuations depend only weakly on $\eta$ for $L=128$, despite the differences observed at larger wave vectors.

\subsection{Characterization of Fractal Dimension}

\begin{figure}[t]
	\centering
	\includegraphics[width=0.8\linewidth]{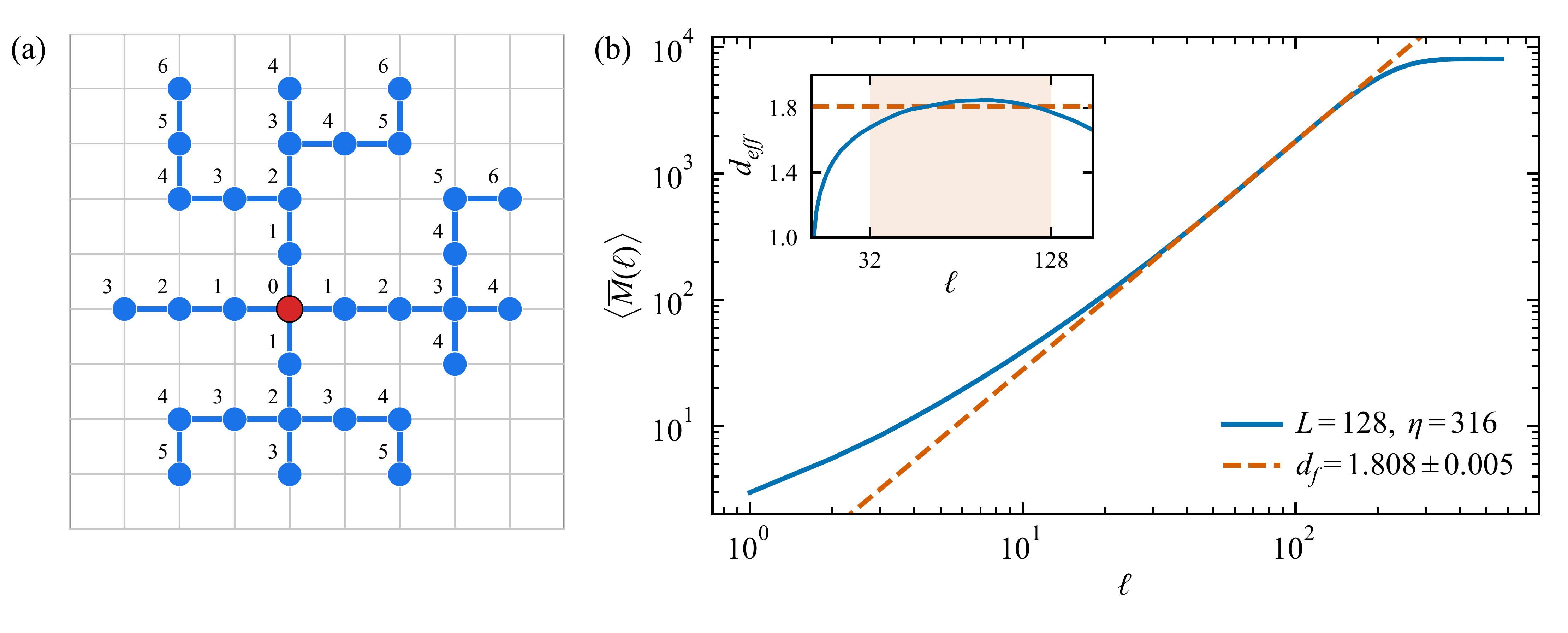}
	\caption{\textbf{Chemical-metric fractal-dimension analysis.} (a) Schematic illustration of the chemical distance $\ell$, defined as the minimum number of nearest-neighbor bonds connecting a site to the red origin along the cluster. (b) Ensemble-averaged cumulative mass $\langle\overline{M}(\ell)\rangle$ versus $\ell$ for $L=128$ and $\eta=316$. For each trajectory, $M(\ell)$ is averaged over multiple occupied origins (overbar), followed by an average over $128$ independent steady-state trajectories (angular brackets). The dashed line is a fit $\langle\overline{M}(\ell)\rangle\sim\ell^{d_f}$, yielding $d_f=1.808\pm0.005$. The inset shows the local effective exponent $d_{\mathrm{eff}}(\ell)$, with the shaded region marking the fitting window.
	}
	\label{figs4}
\end{figure}

To characterize the internal connectivity of the steady-state cluster, we use the cluster-growing method employed for quantum slush~\cite{zhang2024Quantum}, measuring the fractal dimension $d_f$ in the chemical metric~\cite{vicsek1992fractal,Fractal_Mandelbrot}. For the single-seed dynamics under the hard single-NN facilitation constraint considered here, the occupied sites form a single nearest-neighbor-connected cluster. The chemical distance $\ell$ between two occupied sites is the minimum number of nearest-neighbor bonds connecting them along the cluster. As illustrated in Fig.~\ref{figs4}(a), integer values denote the chemical distances measured from a chosen origin (highlighted in red). For a given origin, the cumulative mass $M(\ell)$ represents the total number of occupied sites within a chemical distance $\ell$. We sample multiple occupied sites as origins within each trajectory, denoting the resulting origin-averaged mass as $\overline{M}(\ell)$. Subsequent averaging over independent steady-state trajectories yields the ensemble average $\langle\overline{M}(\ell)\rangle$. We define the fractal dimension $d_f$ through the intermediate-scale scaling relation:
\begin{equation}
\langle\overline{M}(\ell)\rangle \sim \ell^{d_f}.
\end{equation}
We fit the intermediate-distance regime, excluding short-distance lattice effects and large-distance finite-size saturation. The fitting window is selected where the local effective exponent
\begin{equation}
d_{\mathrm{eff}}(\ell) = \frac{d\log \langle\overline{M}(\ell)\rangle}{d\log \ell}
\end{equation}
varies relatively slowly. For $L=128$ and $\eta=316$, $d_{\mathrm{eff}}(\ell)$ remains near $1.8$ within the shaded window in the inset of Fig.~\ref{figs4}(b). A power-law fit over this interval yields
\begin{equation}
d_f = 1.808 \pm 0.005,
\end{equation}
where the reported uncertainty reflects statistical variations across the $128$ independent steady-state trajectories. This exponent quantifies the internal connectivity of the ramified clusters observed in the steady-state configurations
through the scaling of cluster mass with chemical distance.

\end{document}